\documentclass[a4paper,fleqn,usenatbib]{mnras}

\usepackage{amsopn}
\usepackage{hyperref}
\usepackage{xurl}
\usepackage{breakurl}
\usepackage{color}
\usepackage{caption}
\usepackage{relsize}
\usepackage{graphicx}
\usepackage{amsmath}
\usepackage{amssymb}
\usepackage{pdflscape}
\usepackage{rotating}
\usepackage{longtable}
\usepackage{afterpage}

\title[Detecting IMBHs in midiquasars]{Detecting intermediate mass black holes in midiquasars with current and future surveys}
\author[Liodakis, I.]
{I. Liodakis$^{1}$\thanks{yannis.liodakis@utu.fi}\\
$^{1}$Finnish Centre for Astronomy with ESO, University of Turku, Vesilinnantie 5, Turku, FI 20014, Finland\\
}

\begin{document}

\maketitle
\label{firstpage}
\begin{abstract}
The lack of detected intermediate mass black holes poses a gap in our understanding of the growth and evolution of the most exotic of astrophysical objects. Here we investigate the possibility of low-luminosity relativistic jets launched by intermediate mass black holes in the centers of dwarf galaxies. We built population models that allow us to make predictions for their radio emission and quantify their detectability by current and future surveys. We find that the upcoming instruments in optical and radio like the SKA, ngVLA and the Vera C. Rubin Observatory will likely be able to detect a significant fraction ($>38\%$) of such sources population if they exist. In addition, our results suggest that it is not unlikely a small number of midiquasars possibly masquerading as low-luminosity active galactic nuclei may have already been detected by existing surveys.
\end{abstract}


\section{Introduction}\label{introduc}

Since the first observational evidence for black holes (BH) there has been a tremendous leap in our understanding of compact objects and matter under extreme conditions. While the frontier has been pushed in both the low mass range (e.g., \citealp{Fender2006}) as well as the very high mass range (e.g., \citealp{Blandford2019}), the intermediate mass  range  ($10^2$-$10^5$ $M_\odot$ BH, hereafter IMBH) has been elusive. Candidate sources for IMBHs have been ultraluminous X-ray sources, globular clusters and dwarf elliptical galaxies (e.g., \citealp{Mezcua2017}). For far, no confident detection has been made although several candidates exist \citep{Koliopanos2017,Mezcua2017,Greene2019,Mezcua2020}. In this work we focus on the extragalactic IMBH population. Extragalactic IMBH are tremendously important in our understanding of BH grow at very high redshift and the co-evolution with the host galaxy. First indication of IMBHs in the center of low-mass galaxies came from POX 52 \citep{Kunth1987} and NGC 4395 \citep{Filippenko1989} with now more than a few hundred candidate sources. This is naturally expected from the famous $\rm M_{BH}-\sigma$ relation that appears to hold even as far down to $10^5~M_\odot$ BHs (see \citealp[and references therein]{Kormendy2013,Greene2019}). Over the years there is an increasing consensus that such IMBH could be forming active galactic nuclei (AGN) that would be energetic enough to be detected through their optical variability even with a significant star-forming component (e.g., \citealp{Baldassare2018,Baldassare2020}). Judging from both their lighter and heavier relatives it is then natural to assume that a fraction of those systems will form relativistic jets (about 10\% of supermassive BH AGN power jets, \citealp{Wilson1995}). Recent studies have also found radio emission from both nuclear and off-nuclear candidate IMBHs \citep{Davis2020,Reines2020}. In a few cases, studies have found evidence for jets emanating from such sources (e.g., \citealp{Wrobel2006,Yang2020}). It is thus highly likely that a population of IMBH in dwarf galaxies with relativistic jets exists. Given that stellar mass BHs with jets are often referred to as microquasars, we refer to IMBHs with jets as midiquasars. A fraction of those would also be oriented towards our line of sight forming midiblazars. Such a population would be tremendously important given the unique properties of the jets (radio-to-high energy emission, highly polarized emission etc.) that would offer new avenues for detecting IMBHs with current and upcoming surveys.

In \cite{Liodakis2017-III} we found a strong relation between the relativistic beaming-corrected broad-band radio luminosity of the jets in blazars and the mass of the central BH. Using microquasars we found that the extrapolation of the best-fit relation from blazars matches the expectations from stellar-mass BHs well, thus demonstrating that jets are scale invariant.  In this work, we use this universal relation and results from jetted AGN to construct population models in order to make predictions for the possible multiwavelength emission of a population of midiquasars and evaluate their detectability by current and future radio and optical surveys. In section \ref{sec:pop_mod} we describe the population model. In section \ref{sec:multi} we use the results from this work and scaling relations to construct the multiwavelength view of midiquasars and in section \ref{sdss} assess the possibility of finding midiquasars in existing surveys. We summarize our results in section \ref{sec:disc}. We have assumed a flat $\Lambda$CDM Universe with H$_0$=71~km/s/Mpc and $\Omega_m$=0.27 \citep{Komatsu2009}.

\section{Population model}\label{sec:pop_mod}
\begin{figure}
\resizebox{\hsize}{!}{\includegraphics[scale=1]{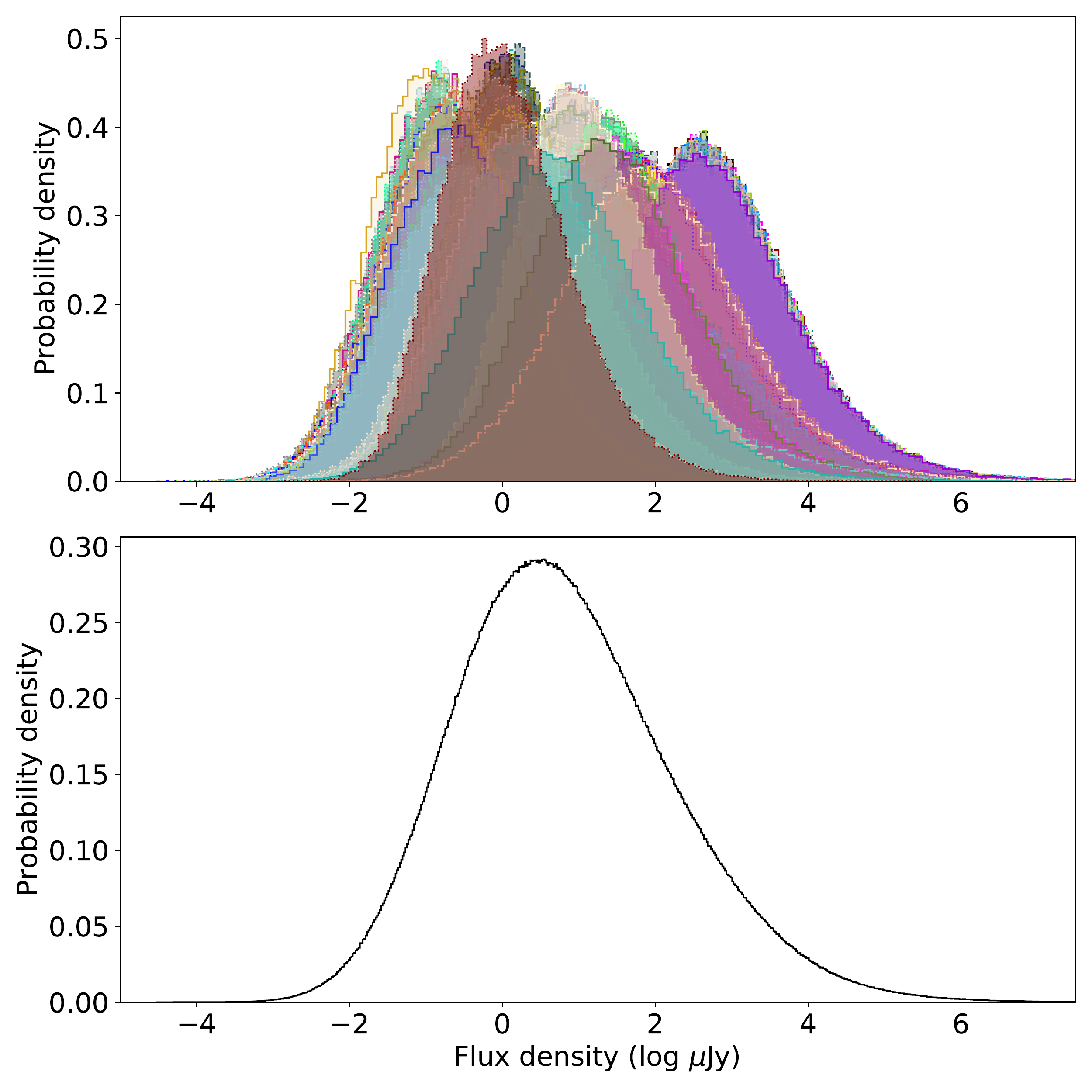} }
 \caption{Top panel: Distribution of the logarithm of the 4.8~GHz flux density for 192 different midiquasar populations. Bottom panel: Combined flux density distribution of all population models. In both panels the bin size was selected following Scott's rule \citep{Scott1979}}
\label{plt:pred}
\end{figure}

Motivated by the similarity of BH systems across the mass range we build our population model similar to \cite{Liodakis2015,Liodakis2017-IV}. However, since there are no available data to optimize the population models, we instead consider different possible scenarios that result in several realizations of a possible midiquasar population. 

We start assuming that the jets are oriented towards the observer randomly and uniformly ($\cos\theta=[0,1]$). To estimate a radio flux-density we require an estimate for the BH mass, a spectral index, distance to the source, and the velocity of the jet. We consider the following scenarios:
\begin{itemize}
\item For the BH mass we consider four possibilities. The masses are: uniformly distributed [$10^2-10^5$] $M_\odot$ ; normally distributed with mean $\mu=4.5\times10^4$ and standard deviation $\sigma=10^4$; log-normally distributed with $\mu=3.6$ and $\sigma=0.35$; following the BH mass distribution of blazars \citep{Liodakis2020} logarithmically shifted to lower values to preserve the shape of the distribution. The range of parameters was chosen such that the resulting extrema of the distributions would lie within the $10^2-10^5$ range.

\item For the spectral index we assume the sources are either in the optically thin or thick regime.  Assuming the intrinsic spectral index distribution to be a Gaussian, we use data from \cite{Hovatta2014} to estimate the mean and standard deviation of said distribution taking into account the errors of each measurement. Following  \cite{Venters2007,Liodakis2017-IV,Liodakis2017-V} those are estimated as,
\begin{equation}
<s>=\frac{\sum^N_{j=1}s_j}{N},
\label{spid_mean}
\end{equation}
\begin{equation} 
<\sigma>=\sqrt{\frac{\sum^N_{j=1}(s_j-<s>)^2}{N}-\sigma_j^2},
\label{spid_std}
\end{equation}
where N is the number of observations, $<s>$ is the intrinsic mean value of the distribution, $<\sigma>$ is the intrinsic standard deviation, and $s_j$, $\sigma_j$ are the observed data and corresponding uncertainty. For the optically thick case we find $<s>=0.21$ and $<\sigma>=0.34$ and for the optically thin $<s>=-1.01$ and $<\sigma>=0.36$.

\item The redshift distribution of the population is assumed to either following the redshift distribution of blazars $\rm z=[0,6.8]$ or their sub-populations: BL Lacs $\rm z=[0,4.4]$ and Flat Spectrum Radio Quasars (FSRQs) $\rm z=[0,6.8]$. Note than the redshift distribution of BL Lacs is considerably skewed towards lower values compared to FSRQs. Redshift estimates for blazars are taken from the BZCAT\footnote{https://www.asdc.asi.it/bzcat/} \citep{Massaro2015}. We additionally consider a scenario of a uniform distribution up to $\rm z=1$, and then exponentially decaying up to $\rm z\approx6.8$ (the maximum of the blazar distribution). The upper limit of the uniform distribution does have an effect on the resulting population. We have tested a simple uniform distribution up to $\rm z=1$, $\rm z=3$, and $\rm z=6.8$ for different simulated populations. As expected a higher $\rm z_{max}$ introduces more fainter sources, which results in a $\sim5-15\%$ decrease in the percentage of detectable sources in different surveys (see below).

\item The velocity of the jets (quantified using the Lorentz factor $\Gamma=1/\sqrt{1-\beta^2}$, $\beta=u/c$) is assumed to either follow the distribution of all blazars ($\Gamma=[1,95]$), BL Lacs ($\Gamma=[1,36]$), FSRQs ($\Gamma=[1,95]$), or Radio galaxies ($\Gamma=[1,36]$) from \cite{Liodakis2018-II}. Radio galaxies have typically slower jets, followed by BL Lacs and then FSRQs. We additionally consider power-law distributions from the population modeling results in \cite{Liodakis2017-IV} for BL Lacs  (P($\Gamma$)$\propto\Gamma^{0.68}$, $\Gamma=[1,16]$) and FSRQs (P($\Gamma$)$\propto\Gamma^{0.50}$, $\Gamma=[1,26]$).
\end{itemize}
The above considerations result in 192 different realizations of a potential midiquasar population. Our simulations consist of the following steps.
First, we draw a random value for the viewing angle, spectral index, redshift, Lorentz factor, and BH mass. The viewing angle and Lorentz factor are   combined to estimate the jet's Doppler factor as $\delta=1/\Gamma(1-\beta\cos\theta)$. Using the BH mass and scaling relation $\rm \log{L}(W/Hz)=(0.83\pm0.17)\times \log(M_{BH}/10^8) + (24.8\pm0.15)$ from \cite{Liodakis2017-III} we estimate the jet's intrinsic  monochromatic luminosity at 4.8~GHz. We have estimated the intrinsic scatter in that relation to be 0.45~dex which we take into account through random sampling. We estimate the observed monochromatic flux density using,
\begin{equation}\label{flux}
S_\nu=\frac{L_\nu D^p}{4\pi{d_L^2}}(1+z)^{1+s},
\end{equation}
where $\rm s$ is the spectral index, and $\rm p=2-s$. We repeat the process $10^5$ times for each of the 192 different midiquasar populations. However, this gives an estimate of the radio emission from the core of the jet. Results from the GLEAM, AT20G and TGSS surveys indicate that core and radio lobes have on average similar contributions to the emission at 150~MHz ($\rm S_{core,150~MHz}/S_{lobe,150~MHz}\sim{1}$) with some scatter \citep{Fan2018,dantonio2019}. Recent results suggest a mean and standard deviation $\mu=0.83\pm0.17$ for $\gamma$-ray loud and $\mu=0.72\pm0.05$ for non-$\gamma$-ray sources \citep{Mooney2021}. We use the average of the two estimates, $\mu=0.775\pm0.17$. To account for the contribution from the radio lobes, we first estimate the unbeamed flux-density of the core at 150~MHz using the above spectral index. Then, we draw a random value for $\rm S_{core,150~MHz}/S_{lobe,150~MHz}$ from a Gaussian distribution with the aforementioned mean and standard deviation. We use that value and the optical thin spectral index $\propto\nu^{-0.7}$ to estimate $\rm S_{lobe}$ at 4.8~GHz without correcting for relativistic effects. The final estimate is the sum of core and lobe flux densities. The flux density distributions for the different populations are shown in Fig. \ref{plt:pred}. We find the median flux density for all the models to be around 4~$\rm\mu{Jy}$ ranging from $\rm\sim0.1~\mu{Jy}$ to $\rm\sim500~\mu{Jy}$. Some models produce simulated sources with flux densities $\rm>1~mJy$ which would make them easily detectable with current instruments, however, those are typically below a few percent for the majority of the individual populations (median of all populations about 3\%) up to $\sim$30\% in the most optimistic models. These are typically high mass ($\rm \sim10^4~M_\odot$), highly beamed ($\rm\delta>10$), and low-z ($\rm z<1.5$) sources. If they exist, they would most likely resemble low luminosity BL Lac objects.

\section{Multiwavelength view of midi-quasars}\label{sec:multi}
\begin{figure}
\resizebox{\hsize}{!}{\includegraphics[scale=1]{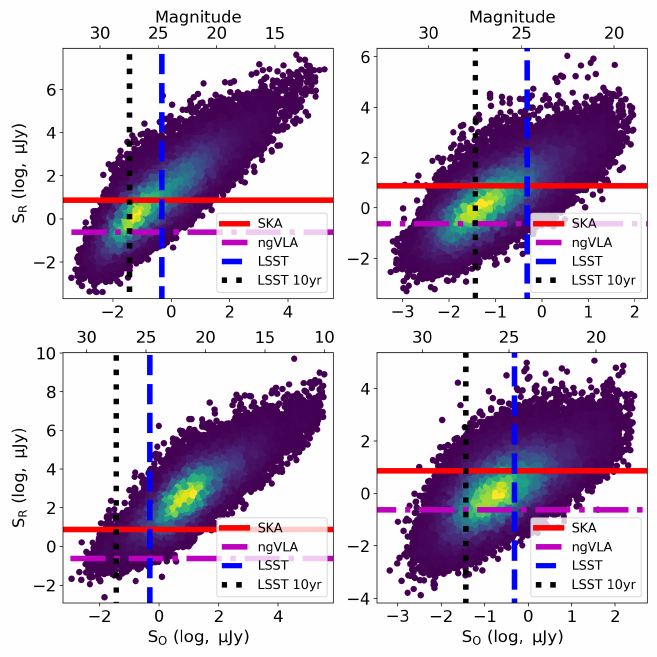} }
 \caption{Comparison between radio and optical flux density (log, $\mu$Jy) for four randomly selected midiquasar populations. The vertical and horizontal lines show the sensitivity limits for SKA (solid red), ngVLA (dash-dotted magenta), LSST snapshot (dashed blue), and the co-added LSST 10 yr (dotted black). The top x-axis shows the apparent magnitude in the R-band assuming the AB magnitude system.}
\label{plt:multi}
\end{figure}

Future surveys in radio (e.g., Square Kilometer Array -- SKA) and optical (e.g., Vera C. Rubin Observatory's Legacy Survey of Space and Time -- LSST) will provide unprecedented opportunities to detect midiquasars. We use the results from the population models and BH mass - host galaxy scaling relations in optical to paint the multiwavelength picture of midiquasars. While the scaling relations are not optimize using IMBHs, they can still provide a useful glimpse of the expectations from the emission of such sources. We estimate the absolute magnitude in the R-band ($\rm M_R$) using  $\rm {\log}M_{BH}= -(0.78 \pm 0.10)M_R -10.39\pm 2.35$ \citep{Wu2002}, which we then convert to apparent magnitude and flux density. Figure \ref{plt:multi} shows a comparison between radio and optical for four randomly selected midiquasar populations (Fig. \ref{plt:multi}), while the different lines show the sensitivity limits for LSST (single exposure -- 24.7$^m$ and 10~yr co-added -- 27.5$^m$\footnote{\url{https://www.lsst.org/scientists/keynumbers}}), ngVLA (0.239 $\rm\mu{Jy}$\footnote{\url{http://library.nrao.edu/public/memos/ngvla/NGVLA_21.pdf}}), and SKA (7.5$\rm\mu{Jy}$, SKA1 band 5 (4-13~GHz)\footnote{\url{https://www.skatelescope.org/wp-content/uploads/2020/02/ScienceCase_band6_Feb2020.pdf}}). Based on these sensitivity limits we can  estimate the fraction of sources in individual populations that could potentially be detected. In the most optimistic cases, SKA will be able to detect $\sim$96.4\% of sources. Similar results are found for LSST, while ngVLA will be able to detect almost all the sources ($\sim$95\% and $\sim$99\% respectively). On average, those percentages are $\sim$40\% for SKA, $\sim88\%$ for ngVLA, and $\sim38\%$ for LSST. On the most pessimistic scenarios $\sim4\%$ for SKA, $39\%$ for ngVLA and $\sim10\%$ LSST. The estimates for the co-added 10-years of LSST observations are about 99\%, 92\% and 64\% for the optimistic, on average, and pessimistic scenarios respectively. Focusing on sources in the [$10^4-10^5$] $M_\odot$ range, which are more likely to be detected in the early years of the upcoming surveys, we only find a 5-15\% improvement in the detectability percentages suggesting that the high-mass end of the midiquasar population will most likely be accessible to the next generation surveys.  It is possible that the true midiquasar population will be a combination of the different scenarios considered above similar, for example, to radio galaxies that show both flat and steep spectrum sources. If we consider the joined model distribution (e.g., Fig. \ref{plt:pred} -- bottom panel) as the true midiquasar population we find that $\sim45\%$ of sources will be detectable by SKA, $85\%$ by ngVLA and $\sim45\%$ by LSST. Detecting midiquasars with instruments in both optical and radio is, of course, limited by the least sensitive of the two. However, the synergy of multiple surveys might prove invaluable to differentiate midiquasars from other low-luminosity objects.

\section{Midiquasars in existing surveys?}\label{sdss}
\begin{figure}
\resizebox{\hsize}{!}{\includegraphics[scale=1]{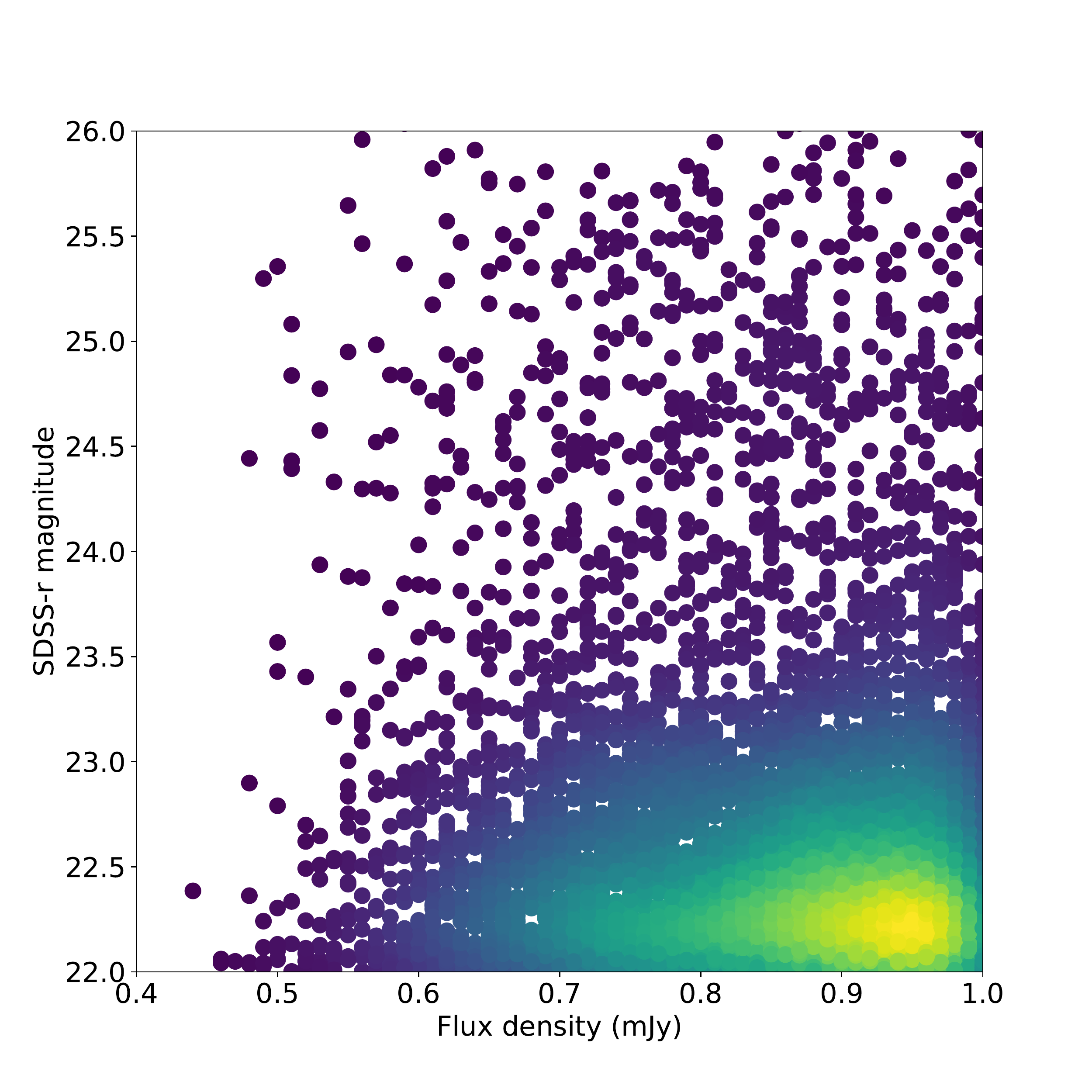} }
 \caption{Comparison between radio flux density (mJy) and optical r-band magnitude for sources in the unified radio catalog without a spectroscopic classification. The color shows the density of observations.}
\label{plt:sdss}
\end{figure}

Our models show a fraction of simulated sources that could be in the detectable range by current surveys. To access that possibility we use the FIRST-NVSS-WENSS-GB6-SDSS Radio Object Catalog \citep{Kimball2008}\footnote{\url{http://www.aoc.nrao.edu/~akimball/radiocat_1.1.shtml}}. From the catalog we select FIRST and NVSS sources that are both the nearest and brightest match to an SDSS counterpart. We use the spectroscopic flags -1 and 0 to select sources that were deemed either unclassifiable or without a spectroscopic match. This allows us to exclude already known sources such as low luminosity stars, high-z quasars, and galaxies. Dwarf galaxies with AGN activity have been found to show atypical spectral features than higher mass AGN \citep{Baldassare2020,Ward2021}. Therefore, our criteria allow us to produce a sample 789,372 ``unidentified/unclassified''  sources. We further limit our sample to sources fainter than 22$^m$ (95\% completeness in SDSS) in the r-band and fainter than 1~mJy (detection threshold for FIRST, \citealp{Steward2018}). Our final list consists of 6013 unidentified/unclassified sources. Figure \ref{plt:sdss} shows the radio flux density and optical magnitude for our unidentified/unclassified sample. We estimate the fraction of simulated sources within the minimum and maximum values for the candidate sample ([22,26] magnitude for the optical and [0.44,1]~mJy for radio) to range from 0.1\% to 11\% with on average 4\%. This exercise is only intended to demonstrate that ``unidentified/unclassified'' sources in existing surveys lie in a similar range as a fraction of the midiquasar populations. The uncertainty in the black hole mass function \citep{Greene2019} and the expected fraction of dwarf galaxies with AGN activity prevents us from converting these percentages to number of sources. Extending the black hole mass function to $<10^6~M_\odot$  will let us further constrain the population models. The majority of the sample is most likely low-luminosity distant AGN and star-forming galaxies, although at the $\sim$mJy level star formation will most likely produce detectable spectral lines \citep{Kouroumpatzakis2021}. Variability, polarization, as well as other traces for star-formation can help exclude such sources as candidates. However, distant AGN will produce very similar signatures, requiring additional tracers and exercising caution.

\section{Summary \& Conclusions}\label{sec:disc}

Motivated by the often found universal scaling relations and the similarity between low- and high-mass black hole systems we have constructed 192 midiquasar population models taking into account possible different realizations of such a population. These models allow us to explore the detectability of midiquasars using upcoming surveys. We find that the  Vera C. Rubin Observatory will be able to detect a significant fraction ($\sim38\%$) of the midiquasar population. SKA and ngVLA are likely to detect about 40\% and 88\% respectively.  The fact that we find a large range of detectability estimates for the different populations can be used to constrain the different possibilities considered above (\S \ref{sec:pop_mod}). While ngVLA appears to outperform SKA, the much larger survey area of SKA will more than compensate for the difference in detection threshold. The SKA will also benefit from its polarization capabilities. The highly polarized nature of synchrotron radiation from the jets will be a clear smoking gun for the presence of jets in these sources (e.g., \citealp{Mandarakas2019,Liodakis2019-II}). In addition, the predictions for the optical emission come from the stellar light of the host galaxy. In low-luminosity and nearby sources the host galaxy can have a significant contribution to the optical-IR emission. This is often true even for jetted AGN (e.g., \citealp{Nilsson2007}). Whether the accretion disk/jet emission will dominate over the host galaxy light or be a subdominant component will likely depend on external factors like the accretion rate. Therefore the predictions for the Vera C. Rubin Observatory should be considered as lower limits. This would then suggest that variability and polarization will likely have an important role in differentiating midiquasars from other low-luminosity sources (e.g., star-forming galaxies).

We additionally evaluated the presence of midiquasars in existing surveys using the FIRST-NVSS-WENSS-GB6-SDSS Radio Object Catalog \citep{Kimball2008}. We identify 6013/789,372 ``unidentified/unclassified''  sources fainter that 1~mJy for radio and 22$^m$ in optical. We estimate that a small fraction (4\% on average) of midiquasars resides within the narrow observed flux-density space of the unidentified sources. While the fraction of sources appears small, that observed range of flux-densities is by no means uncommon. Current radio surveys like the VLASS\footnote{https://science.nrao.edu/science/surveys/vlass} can reach an order of magnitude lower sensitivity  (120~$\rm\mu{Jy}$, \citealp{Lacy2020}). In addition, several optical surveys like the Hyper Suprime-cam wide field survey \citep{Aihara2019} and the Dark Energy Survey \citep{DESCollaboration2021} can reach depths of $26.4^m$ and $24.4^m$ respectively. The above exercise demonstrates that unidentified sources in existing surveys share common optical/radio luminosity space with a fraction of the midiquasar populations. Therefore, it is not impossible that midiquasars have already been detected, but have either gone unnoticed, or masquerading as low luminosity AGN. Mining of existing survey data is highly encouraged.

\section*{Acknowledgments}
We thank the anonymous referee for comments that helped improve this work as well as K. Kouroumpatzakis and A. Zezas for useful discussions. I.L thanks the Institute of Astrophysics - FORTH, at the University of Crete for their hospitality during which this paper was written.

\section*{Data Availability}

The data underlying this article will be shared on reasonable request to the corresponding author.

\bibliographystyle{mnras}
% Use the LaTeX power, use bibtex properly.
\bibliography{bibliography} %graphy.bib}%,bibliography_export.bib}

\label{lastpage}
\end{document}